\documentclass[12pt,a4paper]{article}
\usepackage{amsmath}
\usepackage[dvips]{graphicx}
\input epsf
\usepackage{times}
\begin{document}
\begin{titlepage}
\title{Soft dynamics of large transverse momentum processes}
\author{S.M. Troshin
\\[1ex]
\small  \it Institute for High Energy Physics,\\
\small  \it Protvino, Moscow Region, 142281 Russia} \normalsize
\date{}
\maketitle

\begin{abstract}
We discuss the point that the processes with high transverse momenta could have a
 nonperturbative origin. This discussion is motivated by the fact that there are several experimental
indications that such processes can originate from soft dynamics.
The model combined with unitarity and impact parameter picture  provides
simple mechanism for the hadron production and reproduces power like dependence of the differential
cross-sections in such processes at high transverse momenta and exponential behavior at
small values of $p_\perp$.
This model can be used for support of the argument that the soft dynamics can be responsible for
large transverse momenta processes.

\end{abstract}
\end{titlepage}
\setcounter{page}{2}

\section*{Introduction}
It is widely accepted that perturbative QCD calculations of hard hadronic processes
are in good agreement with experimental data on the differential cross--sections
 at high $p_\perp$. This agreement is certainly one of pQCD successes. However,
power-like dependence of the cross-sections can be based on the different ground.
It is well known that such dependence has been obtained \cite{berman,matv,farr} on the basis
 of the parton model and dimensional
scaling prior to the perturbative QCD calculations.

The perturbative QCD is well elaborated
theory and
provides predictions for many phenomena and among them are helicity conservation \cite{lepage}
in hard processes and existence of the new form of matter with free quarks and gluons
-- quark-gluon plasma \cite{coll}.  However, violation
of helicity conservation has been observed in many experiments and has a rather long
 experimental history \cite{hoyer},
while strongly interacting
deconfined  matter in heavy-ion collisions at RHIC was revealed just recently \cite{rhic}.
It appeared that instead of gas of free quarks and gluons the matter created at RHIC is an almost
perfect liquid. The nature of this new form of matter is not known and the variety of models
has been proposed to treat its properties \cite{models}. This important
result is that beyond the critical values of density and temperature
 the matter is strongly correlated
and reveals high degree of the coherence. The elliptic flow and constituent quark scaling of the
observable $v_2$ demonstrated  an importance of the constituent quarks \cite{volosh} and their role as
effective degrees of freedom of the newly discovered form of matter.
Generally speaking this result has shown again an importance of the nonperturbative
effects in the region where such effects were usually not expected.

It appeared also that the inclusive
cross-section of pion production in Au+Au heavy-ion collisions has a power-like tail in the
 high $p_\perp$ region \cite{inclrhic}.  It is no need to comment that such power-like dependence
  always being considered as a manifestation of the hard, short distance
 processes where asymptotic freedom is at work. Thus power-like  behavior of inclusive cross-sections
  and the strongly
 interacting nature of quark matter can be attributed to different dynamics. But it is difficult to
  imagine how both phenomena can coexist in the strongly interacting coherent medium
  observed at RHIC when thermalization occurs at very early stage of reaction.  It seems that it is
 natural to suppose that
 they can have the same origin. One  should  in principle
 arrive then to conclusion
 that the power-like
 dependence of the differential cross-sections should not necessary be associated with the
 processes treated by  perturbative QCD. This viewpoint has   support
 in the results on polarization measurements which also indicate possibility of power-like behavior
 due to soft dynamics \cite{hoyer}. It should be recollected that the energies
 where power-like dependence in exclusive
 processes was experimentally observed are evidently
  too low to be considered as a true asymptotic perturbative QCD
 domain.
Such dependence of the differential cross-section of exclusive processes  also might
 be related to nonperturbative dynamics \cite{usela} because it would be rather strange
 to expect that perturbative QCD can be applied in this region.

 Note also that exponent in the
 inclusive cross-section dependence at high $p_\perp$ varies from 6 to 12 in different kinematic
 regions and in average is almost twice bigger than 4 predicted by the
 leading twist perturbative QCD calculations.

 In this note in framework of the nonperturbative approach and its realization in
  the particular model we show that
 power-like dependence of inclusive cross-section at high $p_\perp$ can result from
  soft interaction dynamics, which produces simultaneously exponential dependence in the region
   of small transverse momenta.

\section{Chiral quark model and an effective field in hadron interaction}

In the nonperturbative sector of QCD the  two important
phenomena,  confinement and spontaneous breaking of chiral symmetry ($\chi$SB)\cite{mnh}
should be reproduced.
The  relevant scales   are characterized by the
parameters $\Lambda _{QCD}$ and $\Lambda _\chi $, respectively.  Chiral $SU(3)_L\times
SU(3)_R$ symmetry is spontaneously  broken  at the distances
in  the range between
these two scales.  The $\chi$SB mechanism leads
to generation of quark masses and appearance of quark condensates. It describes
transition of current into  constituent quarks.
  Constituent quarks are the quasiparticles, i.e. they
are a coherent superposition of bare  quarks, their masses
have a magnitude comparable to  a hadron mass scale.  Therefore
hadron  is often represented as a loosely bounded system of the
constituent quarks.
These observations on the hadron structure lead
to  understanding of several regularities observed in hadron
interactions at large distances. It is well known  that such picture  provides
reasonable  values  for the static characteristics of hadrons, for
 instance, their magnetic moments. The other well known straightforward  result
   is  appearance of the Goldstone bosons.

   We will appeal to the   model for the hadron scattering
 described in \cite{csn} and develop it further for the purposes mentioned above.
It is  based on the ideas of chiral quark models and mean field approximation
to hadron structure (cf. \cite{diak}).
As it was already mentioned constituent quarks and Goldstone bosons are the effective
degrees of freedom in the chiral quark model and therefore we consider a
 hadron consisting of the valence
constituent quarks located in the central core which is embedded into  a quark
condensate. Collective excitations of the condensate are the Goldstone bosons
and the constituent quarks interact via exchange
of Goldstone bosons; this interaction is mainly due to a pion field and is of the flavor--
 and spin--exchange nature. Thus, quarks generate a strong field which
binds them.

The  general form of the effective lagrangian relevant for
description of the non--perturbative phase of QCD proposed in \cite{gold}
is the following
${\cal{L}}_{QCD}\rightarrow {\cal{L}}_{eff}$ and includes the three terms \[
{\cal{L}}_{eff}={\cal{L}}_\chi +{\cal{L}}_I+{\cal{L}}_C.\label{ef} \]
Here ${\cal{L}}_\chi $ is  responsible for the spontaneous
chiral symmetry breaking and turns on first.  To account for the
constituent quark interaction and confinement the terms ${\cal{L}}_I$
and ${\cal{L}}_C$ are introduced.  The  ${\cal{L}}_I$ and
${\cal{L}}_C$ do not affect the internal structure of the constituent
quarks.

The picture of a hadron consisting of constituent quarks embedded
 into quark condensate implies that overlapping and interaction of
peripheral clouds   occur at the first stage of hadron interaction.
At this stage the part of the effective lagrangian ${\cal{L}}_C$ is turned off.
(It turned on again in the final stage of the reaction).
Nonlinear field couplings  could transform then the kinetic energy to
internal energy and mechanism of such transformations was discussed
 by Heisenberg \cite{heis} and  Carruthers \cite{carr}.
As a result massive
virtual quarks appear in the overlapping region and  some effective
field is generated.
Valence constituent quarks  located in the central part of hadrons are
supposed to scatter simultaneously in a quasi-independent way by this effective
 field. Thus, this picture assumes  deconfinement at the initial stage of
 the hadron collisions and  generation of common for both hadrons mean field during this stage.
It might happen
 that the similar mechanism is responsible for the creation of the
  strongly interacting matter in heavy-ion collisions due to strong interaction
  of constituent quarks with chiral field. The
   existence of the massive quark-antiquark matter at the stage
preceding the hadronization seems to be
supported  by the experimental data obtained
at CERN SPS and RHIC and can be reconciled with the possible
flash-like particle emission \cite{jenk}
  (cf. also  \cite{volosh,biro} and references
therein\footnote{It should be noted that the various aspects of similarity between hadron
and heavy-ion interaction dynamics in the historical context of
search for the quark-gluon plasma are discussed in detail in \cite{weiner}}).

 Massive virtual quarks (newly formed constituent quarks) play a role
of scatterers (the part of the effective lagrangian ${\cal{L}}_I$ plays the main role
 in this phase of hadron
 interaction) for the valence quarks in elastic hadron scattering and
 their hadronization leads to
production of the secondary particles in multiparticle processes.
 To estimate number
of such quarks
one could assume that  part of hadron energy carried by
the outer condensate clouds is released in the overlap region
 to generate those quarks. Then this number can be estimated  by:
 \begin{equation} \tilde{N}(s,b)\,\propto
\,\frac{(1-\langle k_Q\rangle)\sqrt{s}}{m_Q}\;D^{h_1}_c\otimes D^{h_2}_c
\equiv N_0(s)D_C(b),
\label{Nsbt}
\end{equation} where $m_Q$ -- constituent quark mass, $\langle k_Q\rangle $ --
average fraction of
hadron  energy carried  by  the constituent valence quarks. Function $D^h_c$
describes condensate distribution inside the hadron $h$, and $b$ is
an impact parameter of the colliding hadrons.

Thus, $\tilde{N}(s,b)$ quarks appear in addition to $N=n_{h_1}+n_{h_2}$
valence quarks. In elastic scattering those quarks are transient
ones: they are transformed back into the condensates of the final
hadrons. Calculation of elastic scattering amplitude has been performed
in \cite{csn}.
However,  valence quarks can excite a part of the cloud of the virtual massive
quarks and these virtual massive
 quarks will subsequently fragment into the multiparticle
final states. The triggering mechanism of such excitation is based on
 the feature of chiral quark model that constituent quark $Q_\uparrow$
with transverse spin in up-direction can fluctuate into Goldstone boson and
  another constituent quark $Q'_\downarrow$ with opposite spin direction,
   i. e. perform a spin-flip transition \cite{cheng} with impact parameter distortion:
\begin{equation}\label{trans}
Q_{\uparrow,\downarrow}\to GB+Q'_{\downarrow,\uparrow}\to Q+\bar Q'+Q'_{\downarrow,\uparrow}.
\end{equation}
It is important to note here that the shift of $\tilde{\bf b}$
(the impact parameter of final particle)
is translated to the shift of the impact parameter of the initial particles according
to the relation between impact parameters in the multiparticle production\cite{webb}:
\begin{equation}\label{bi}
{\bf b}=\sum_i x_i{ \tilde{\bf  b}_i}.
\end{equation}
The variable $\tilde b$ is conjugated to the transverse momentum of final pion.

We consider production of pions  in the fragmentation region, i.e.
at large $x_F$ and therefore use an approximate relation
\begin{equation}\label{bx}
b\simeq x_F\tilde b,
\end{equation}
which results from Eq. (\ref{bi})\footnote{We make here an additional assumption on the
small values of Feynman $x$ for other particles}.

We will discuss how this mechanism can produce power-like dependence of inclusive
cross-section in what follows.
This mechanism is responsible for the  particle production
in the fragmentation region and should lead to  strong correlations between
secondary particles. It means that correlations exist
between particles from the same (short--range correlations)
and different clusters (long--range correlations)
 and, in particular, the forward--backward
multiplicity correlations should be observed. This mechanism can be called
  a correlated cluster
production mechanism.

 Direct, i.e. not induced by the valence
quarks hadronization of  the part of the massive $\tilde{N}(s,b)$
constituent quarks, leads to  formation of the multiparticle
final states in the central region.
These states are also correlated due to strong constituent quark interactions and it would
results in  correlations in the multiplicity
distribution.

\section{Inclusive pion cross-sections at high transverse momenta}

Unpolarized cross-section of pion production with account of unitarity in the
rational form has the following form \cite{tmf}:
\begin{equation}
\frac{d\sigma}{d\xi}= 8\pi\int_0^\infty
bdb\frac{I_0(s,b,\xi)}{|1-iU(s,b)|^2}\label{unp}.
\end{equation}

In the fragmentation region we can simplify the problem and consider
for example
the process of pion-production as a quasi two-particle reaction,
where the second final particle has a mass $M^2\simeq (1-x_F)s$.
 The amplitude of this quasi two-particle reaction in the pure imaginary
 case (which we consider for simplicity) can be written in the form
 \begin{equation}
F(s,p_\perp,x_F)= \frac{is}{x_F^2\pi^2}\int_0^\infty
bdbJ_0(bp_\perp/x_F)\frac{I^{1/2}_0(s,b,x_F)}{1+U(s,b)}\label{amp}.
\end{equation}

We  consider neutral pion production
and take constituent quarks $U$ and $D$ masses to be equal\footnote{Charged pion
 production can be treated in a similar way.}. Pion appear as a result of transition
$(U,D)_{\uparrow,\downarrow}\to (U,D)_{\downarrow,\uparrow}+\pi^0$.
The function $I_0^{\pi0}(s,b,x_F)$ according to quasi-independent nature
of constituent quark-scattering
can be represented then as a product in a way similar to the
 case of elastic scattering \cite{csn}:
\begin{equation}
 I_0^{\pi^0}(s,b,x_F)= \left[\prod^{N-1}_{Q=1} \langle f_Q(s,b)\rangle\right]\langle
f_{\pi^0}(s,b,x_F)\rangle,
\end{equation}
 where $N$ is the total number of quarks in the colliding
hadrons.

In the model the $b$--dependencies of the amplitudes $\langle f_{Q}
\rangle $ and $\langle f_{\pi^0} \rangle $
are related to the strong
formfactor of the constituent quark and transitional spin-flip formfactor
 respectively.
 The strong interaction radius of constituent
quark is determined by its mass. We suppose that the corresponding radius of transitional formfactor
 is determined by the mass ${m}_Q$ and factor $\kappa<1$ (which takes
into account reduction of the radius due to spin flip), i.e.
$r^{flip}_{\pi^0} = \kappa\zeta /{m}_{{Q}}$ and the corresponding function
$f_{\pi^0}(s,b,x_F)$ has the form
\begin{equation}\label{ftr}
f_{\pi^0}(s,b,x_F)=g_{flip}(x_F)\exp\left(-\frac{{m}_Q}{\kappa\zeta}b\right)
\end{equation}

 The expression for
$I_0(s,b,x_F)$ can be rewritten then in the following form:
\begin{equation}\label{iol}
I_0(s,b,x_F) =\frac{\bar{g}(x_F)}{g_Q(s)}
U(s,b)\exp[-\Delta m_Q b/\zeta ],
\end{equation}
where the mass difference $\Delta m_Q\equiv {m}_Q/\kappa-m_Q$ and $\bar{g}(x_F)$ is the function whose
dependence  on Feynman $x_F$ in the model is not fixed.

Now we can consider $p_\perp$- and $x_F$-dependencies
of the pion production
cross-section and we start with angular distribution\footnote{One should remember that
all formulas below are valid for the fragmentation region, i.e. for the large $x_F$ region}.
The corresponding amplitude
$F(s,p_\perp,x_F)$ can be calculated analytically. To do so
 we continue the amplitudes
$F(s,\beta, x_F),\,\beta =b^2$, where
\[
F(s,\beta, x_F)=\frac{1}{x_F^2}\frac{I^{1/2}_0(s,\beta,x_F)}{1+U(s,\beta)}
\]
to the complex
 $\beta $--plane and transform the Fourier--Bessel integral over impact
parameter into the integral in the complex $\beta $ -- plane over
the contour $C$ which goes around the positive semiaxis.
Then  for the  amplitude $F(s,p_\perp,x_F)$ the following
representation takes place:
\begin{equation}\label{impl}
F(s,p_\perp,x_F)  =  -\frac{is}{2\pi ^3}\int_C d\beta
F(s,\beta, x_F)K_0(\sqrt{-p^2_\perp\beta/x_F })
\end{equation}
where $K_0(x)$ is the modified Bessel function. The amplitude
 $F(s,\beta, x_F)$ has the poles in the
$\beta $--plane determined by  equation
\begin{equation}\label{polloc}
\beta _n(s)=\frac{\zeta ^2}{ M^2}\,\left\{\,\ln g(s)+\,i \pi
n\,\right\}^2,\, n=\pm 1, \pm 3,\ldots
\end{equation}
and a  branching point at $\beta =0$.
The poles and cut contributions determine the behavior of
the inclusive cross-section of pion production at moderate and large
values of $p_\perp$ correspondingly, i.e. it will have in
the region of large $p_\perp$ power-like dependence on $p_\perp$:
\begin{equation}\label{dsig}
\frac {d\sigma}{d\xi}\propto G_c^2(s,x_F)(1+\frac{p_\perp^2}{ x^2_F \bar{M}^2})^{-3},
\end{equation}
while at smaller values of $p_\perp$ inclusive cross-section would have the exponential $p_\perp$-dependence:
\begin{equation}\label{dsig1}
\frac {d\sigma}{d\xi}\propto G_p^2(s,x_F)\exp(-\frac{2\pi \zeta }{M}\frac{p_\perp}{x_F}).
\end{equation}
It should be noted that the value of
the inverse slope $Mx_F/2\pi \zeta $ is about a pion mass value since parameter
$\zeta =2 $ is fixed in the model .
The data for the pion production are available at
the moderate values of $p_\perp$ and the experimental fits to the
data \cite{data} of the  form \[
A(1-x_F)^n e^{-B(x_F)p_\perp}\]
 just
follow to Eq. (\ref{dsig1}) when relevant parameterizations for the
function $\bar g(x_F)$\footnote{This function determines $x_F$-dependence of
the functions $G_p$ and $G_c$} are chosen.

The contribution of poles
can be interpreted as a result of a collective resonances excited in the constituent quark matter
in the interaction region (where interaction radius is
$R^2(s)=\frac{\zeta ^2}{ M^2}\,\ln g(s)$). These resonances could have a relation to the collective
resonance phenomena discussed in \cite{goldh}.

At high values of $p_\perp$ the power-like dependence
should take place according to Eq. \ref{dsig}.  In the energy region of $\sqrt{s}\leq 2$ TeV
the functions $G_p$ and $G_c$ have very slow variation with energy due to the numerical values
of parameters \cite{preas}.It should be noted that Eq. (\ref{dsig}) leads to $p_\perp^{-6}$
dependence at large $p_\perp$. This is in  agreement
with $p_\perp^{-N}$ (with the exponent $N=6.2\pm 0.6$) dependence of
the inclusive cross-section of $\pi^0$-production observed
 in forward region at large $p_\perp$ at RHIC \cite{star}. The exponent $N$ does not depend
 on $x_F$ and choosing relevant function $\bar g(x_F)$ the $(1-x_F)^{5.1\pm 0.6}$ dependence
 of experimental data can be easily reproduced. Thus, in the approach with effective degrees of freedom --
 constituent quarks and Goldstone bosons -- differential cross--section at high transverse momenta
 has a typical power-like dependence on $p_\perp$ and is in agreement with experimental data.
It originates from the singularity at zero impact parameter $b=0$ (head-on collision).
In the framework of the model
this singularity should be related to the  constituent quarks location  at the hadron core.

\section*{Conclusion}
The proposed mechanism deals with effective degrees of freedom and takes into
account collective aspects of the nonperturbative QCD dynamics.
We discussed here  particle production in the fragmentation region and have
shown that the power-like behavior of the differential cross-sections at large transverse
momenta can be obtained in the approach which has a nonperturbative origin.
 It might happen that
the transient stage in hadron and heavy-ion interactions have
a common nature and represent a strongly
interacting matter of constituent quarks.
This interaction should then be identical to the one
responsible for  formation of more
 complex multiquark states. Such interaction would result in collective
 rotation of the quark matter as a whole produced in the non-central heavy-ion collisions.
 This rotation should be taken then into account since it will affect the elliptic flow and
 would probably result in strong spin correlations of the produced particles similar to
  correlations predicted for hadron processes \cite{ipl}.

 We would also like to note that
this model explains an exponential dependence of inclusive
cross--section in the region of moderate transverse momenta and
  provides a reasonable description of
the energy dependence of mean multiplicity leading to
its power-like growth with a small exponent \cite{jpg} as
a combined effect of the unitarity and account for the  phase preceding
 hadronization when massive quark--antiquark pairs are generated.

\section*{Acknowledgement}
Author is grateful to N.E. Tyurin for many interesting discussions, comments and
remarks.

\end{document}